# On the possibility of Galactic Cosmic Ray-induced radiolysis-powered life in subsurface environments in the Universe

Dimitra Atri[*]

**Abstract**

Photosynthesis is a mechanism developed by terrestrial life to utilize the energy from photons of solar origin for biological use. Subsurface regions are isolated from the photosphere, and consequently are incapable of utilizing this energy. This opens up the opportunity for life to evolve alternate mechanisms for harvesting available energy. Bacterium *Candidatus* Desulforudis audaxviator, found 2.8 km deep in a South African mine, harvests energy from radiolysis, induced by particles emitted from radioactive U, Th and K present in surrounding rock. Another radiation source in the subsurface environments is secondary particles generated by Galactic Cosmic Rays (GCRs). Using Monte Carlo simulations, it is shown that it is a steady source of energy comparable to that produced by radioactive substances, and the possibility of a slow metabolizing life flourishing on it cannot be ruled out. Two mechanisms are proposed through which GCR-induced secondary particles can be utilized for biological use in subsurface environments: (1) GCRs injecting energy in the environment through particle-induced radiolysis, and (2) organic synthesis from GCR secondaries interacting with the medium. Laboratory experiments to test these hypotheses are also proposed. Implications of these mechanisms on finding life in the Solar System and elsewhere in the Universe are discussed.

## 1. Introduction

Radiation in the form of photons is the primary way by which solar energy is transferred to living systems. The energy of a typical photon is between 2 - 3 eV (Visible Red – 1.8 eV and Visible Blue – 3.1 eV), and an abundant supply of such photons makes it convenient for life to utilize it for metabolic purposes. Even though most of the energy is lost in heat, the overall efficiency of 0.1-8% is sufficient to power metabolism (Zhu et al., 2008). Studies of ionizing radiation, on the other hand have mostly been associated with health-effects on humans (United Nations Report, 2013; Brenner et al., 2003) and studies of radiation resistance on microbes (Mattimore and Battista, 1996; Battista, 1997). By definition, ionizing radiation has the energy to ionize, or to eject at least one electron from a neutral atom or a molecule, which is 13.6 eV for a neutral hydrogen atom, for example. Ionizing radiation can interact with the DNA and cause reparable or irreparable damage, depending on the type and energy of the radiation (Ward, 1988). Such damages have the potential to modify the genetic code through mutations, alter the way the DNA functions, and transfer mutations to the next generation(s) (Dubrova et al., 2000; Dubrova and Plumb, 2002). Radiation damage can also cause cancer as seen in a number of studies with UV and particle interactions with humans, primarily in context of oncological studies, nuclear accidents and astronaut health in outer space (Cucinotta and Durante, 2006; Durante and Cucinotta, 2008, NCRP Report, 2002, 2010). Nevertheless, a high dosage of ionizing radiation can force organisms to develop mechanisms that enable them to survive in extreme conditions (Thornley, 1963).

---

[*] *Blue Marble Space Institute of Science, 1001 4th Ave, Suite 3201, Seattle, WA 98154, USA, dimitra@bmsis.org*

Geochemical processes are well-known to support subsurface life (Ghiorse, 1997; Fernández-Remolar et al., 2008). An important shift in our understanding came about when studies revealed that subsurface life can be independently supported by radiolysis, where the source of radiation is particles emitted from the decay of radioactive substances (Lin et al., 2005; Onstott et al., 2006). Radioactive materials present deep underground produce secondary particles such as alpha, beta and gamma radiation. Secondary particles interact with the environment and provide energy for chemical change. Organisms can use these particles indirectly for metabolic purposes. *Candidatus* Desulforudis audaxviator is such an example, which thrives in an radiolysis-powered ecosystem (Figure 1(a)). Radiation from radioactive rock dissociates $H_2O$ into a number of radicals, useful for biological reactions. It is able to extract carbon from dissolved $CO_2$ and nitrogen in the form of ammonia from the rock, and utilize them to synthesize amino acids. Such an organism can potentially thrive in subsurface environments on Mars, Moon, Europa or other planetary systems in the presence of radioactive substances.

*Ca.* D. audaxviator thrives in a 2.8 km deep South African gold mine (Lin et al., 2006; Chivian et al., 2008). A comprehensive analysis on the availability of nuclear power for deep surface microbes has also been done (Lin et al., 2005). Radiolytic dissociation of water due to radiogenic decay of U, Th and K in rock produces a number of radicals, and generates molecular hydrogen ($H_2$), along with other biologically useful products. Radiolytically generated chemicals provide the necessary energy and nutrients to the system, sulfate ($SO_4^{2-}$) reduction is the dominant electron-accepting process and $H_2$ and formate are primary electron donors. It provides them with the energy to sustain a minimal metabolism. A part of the energy is also utilized to repair damage caused by ionizing radiation. Studies have shown that $H_2$ produced through geochemical processes can also be utilized for metabolism, independent of photosynthesis, using similar mechanisms (Stevens and McKinley, 1995).

In light of these findings, it is reasonable to propose that instead of or in addition to *in situ* energy sources, the radiation (potentially supporting radiolysis-based life) can be of cosmic origin. GCRs are charged particles, mostly protons, originating beyond the solar system (Gaisser, 1990; Dorman, 2004; Stanev, 2010). They have relatively lower flux but possess much higher energy than other radiation sources on Earth, and have noteworthy biological effects on terrestrial life and possibly on extrasolar planets (Dartnell, 2011; Atri et al., 2013; Atri and Melott, 2014). Upon interaction with a planetary atmosphere or surface, GCRs produce a cascade of secondary particles, that include electrons, positrons, gamma rays, neutrons and muons (Gaisser, 1990). Muons can travel several kilometers deep below the surface depending on their energy (Dorman, 2004; Stanev, 2010). These secondary particles can induce radiolysis and as described later, possibly power a subsurface ecosystem.

2. Proposed Mechanisms

2.1. Galactic Cosmic Ray-induced radiolysis

As discussed in the previous section, radioactive rocks provide an *in situ* production source of energy for subsurface life. A similar mechanism is proposed in this section. However, the energy source considered here is extraterrestrial. For planets with substantial atmospheres, the primary GCR particles strike the atmosphere and produce a cascade of secondary particles, also referred to as an air shower (Gaisser, 1990). If a planet lacks an atmosphere, particles are able to directly strike the surface and the cascade of secondary particles is able to propagate underground (Mei and Hime, 2006). The secondary particles produced in the cascade, such as pions and kaons, are highly unstable, and quickly decay to other particles including beta particles (electrons and positrons) and gamma-rays. It must be noted that beta particles and gamma-rays are also produced during radioactive decay in underground rock. When charged pions decay muons are produced, and these can travel several kilometers deep depending on their energy. It should be emphasized that all types of charged particles such as electrons and positrons can contribute to radiolysis in regions near the surface, but muons become important at greater depths where the flux of other particles becomes nearly zero. They only undergo electromagnetic interactions and lose 2-8 MeV energy per gram per square cm in the material they traverse (Groom et al., 2001). The energy loss is a combination of ionization, bremsstrahlung, pair production and inelastic nuclear scattering (Heisinger et al., 2002).

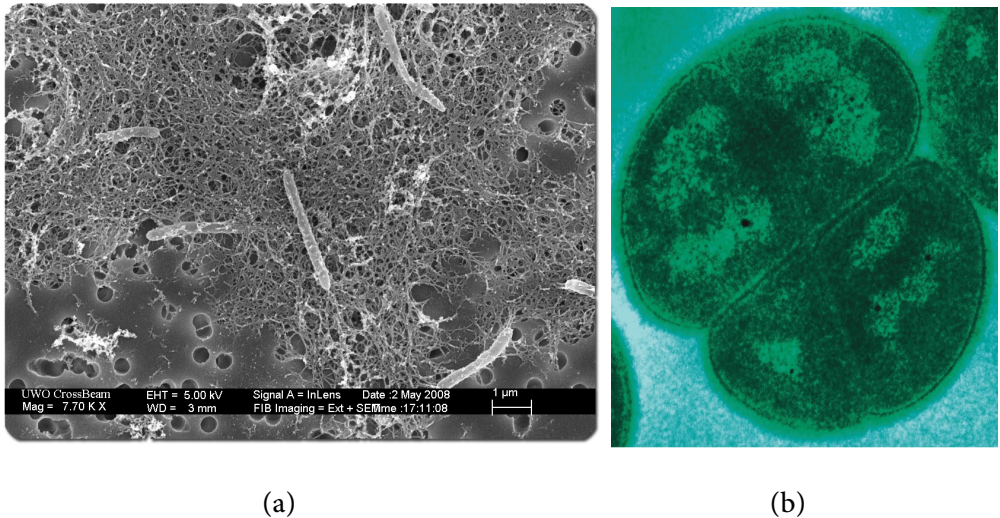

(a)                                                              (b)

Figure 1: (a) A colony of *Ca.* D. audaxviator, discovered in a 2.8 km deep gold mine near Johannesburg, South Africa.
https://upload.wikimedia.org/wikipedia/commons/1/1e/Desulforudis_audaxviator.jpg [Public domain], via Wikimedia Commons.
 (b) TEM of *D. radiodurans* acquired in the laboratory of Michael Daly, Uniformed Services University, Bethesda, MD, USA.
http://www.usuhs.mil/pat/deinococcus/index_20.htm [Public domain], via Wikimedia Commons.

Energetic particles are capable of destroying organic materials present on the surface or subsurface environment of planets with thin atmospheres. A number of studies have been devoted on studying the impact of GCRs on the possible destruction of organic matter on Mars (Pavlov et al., 2012; Pavlov et al., 2002; Dartnell et al., 2007). Mars' surface radiation dose is several orders of magnitude higher than that of the Earth due to its thin atmosphere. Exposure to high environmental radiation dose also leads to substantial energy expenditure in repairing radiation-induced damage. Bacterium *Deinococcus radiodurans* is the focus of numerous studies and is the most radioresistant microbe with a LD10 at 15 kGy (Figure 1(b)). *D. radiodurans* has a specialized recombination system for DNA repair from ionizing radiation. Interestingly, many fungal species also have very high radiation resistance. Many fungi have 10% survival chance or LD10 values exceeding 5 kGy. Other species, such as *Ustilago maydis* are also known to have extreme radiation resistance but have a different mechanism to cope with compared to *D. radiodurans*. For comparison, mammals have very low radiation resistance and a dose between 5 to 10 Gy is lethal. It has been found that the action of proteins is primarily responsible for the repair mechanism (Holloman et al., 2007). An abundance of highly melanized fungal spores in the early Cretaceous period deposits has been uncovered, where other plant and animal species have died out (Hulot and Gallet, 2003). These types of fungi can be found in high-altitude terrains, Arctic and Antarctic regions, and in the Evolution Canyon in Israel.

Radiotropism is a term used to describe the growth of fungi from exposure to radionuclides (Zhdanova et al., 2004). Most radionuclides emit beta (electrons and positrons) and gamma (photons) radiation and studies have shown that exposure to both sources promoted directional growth of fungi (Zhdanova et al., 2004). Exposure to $^{121}$Sn and $^{137}$Cs sources showed that the spore germination in species increased considerably, a phenomenon also known as 'radiostimulation' (Zhdanova et al., 2004). Presence of several microorganisms has been studied in the Russian Mir Space Station as well as the International Space Station (Alekhova et al., 2005). The organisms consist of both bacteria and fungi, and are exposed to about 4 cGy of ionizing radiation per year. Many bacteria and fungi in these environments have been found to be pigmented or melanized (Baranov et al., 2006). Dadachova et al. concluded that ionizing radiation changed the electron spin resonance (ESR) signal of melanin, making it more efficient in acting as an ionizing radiation transducer (Dadachova et al., 2007). Experiments have shown increased metabolic activity in melanized *Cryptococcus neoformans* relative to non-melanized ones, indicating the enhancement of the electron transfer properties of melanin. The authors concluded that melanin played a major role in protecting the organism against ionizing radiation and these radioprotective properties arise due to its chemical composition, free radical quenching, and spherical spatial arrangement (Dadachova et al., 2008).

An abundance of highly melanized fungal spores in the early Cretaceous period deposits has been uncovered, where other plant and animal species have died out (Hulot and Gallet, 2003). These types of fungi can be found in high-altitude terrains, Arctic and Antarctic regions, and in the Evolution Canyon in Israel. The south-facing slope of the canyon receives 2-8 times higher solar radiation than those on the north, and *Aspergillus niger* found there contains 300% higher levels of melanin than that found in the north facing slope (Singaravelan et al., 2008). Several groups have studied these effects on melanized fungal species at Chernobyl accident site (Mironenko et al., 2000) and in nuclear reactor pool water (Mal'tsev et al, 1996).

The discussion above shows that organisms can develop mechanisms to thrive and to repair damage under the exposure of high radiation dose and in one case, harvest it for metabolic purposes, but it remains to be shown that the ionizing energy produced by secondary particles is comparable to energies observed in radioactive decay processes known to support radiolysis-based life.

The GEANT4 package was used to model the energy deposition rate in subsurface environments. It is a widely used package and is considered a gold standard in modeling particle interactions. The code models all particle interactions and it traverses through a medium and has been well tested in the planetary science community. As the particles traverse in the subsurface medium, they lose energy by ionizing it. At greater depths, muons become important because of their long range, as discussed earlier. Muons transfer their large kinetic energy to the medium, and also form Muonium ($\mu^+e^-$), which has been found useful for various chemical and biological reactions due to its similarities with the hydrogen atom (Percival et al., 1978). It should be noted that below 15 km, the energy deposition rate is constant due to contribution from neutrinos.

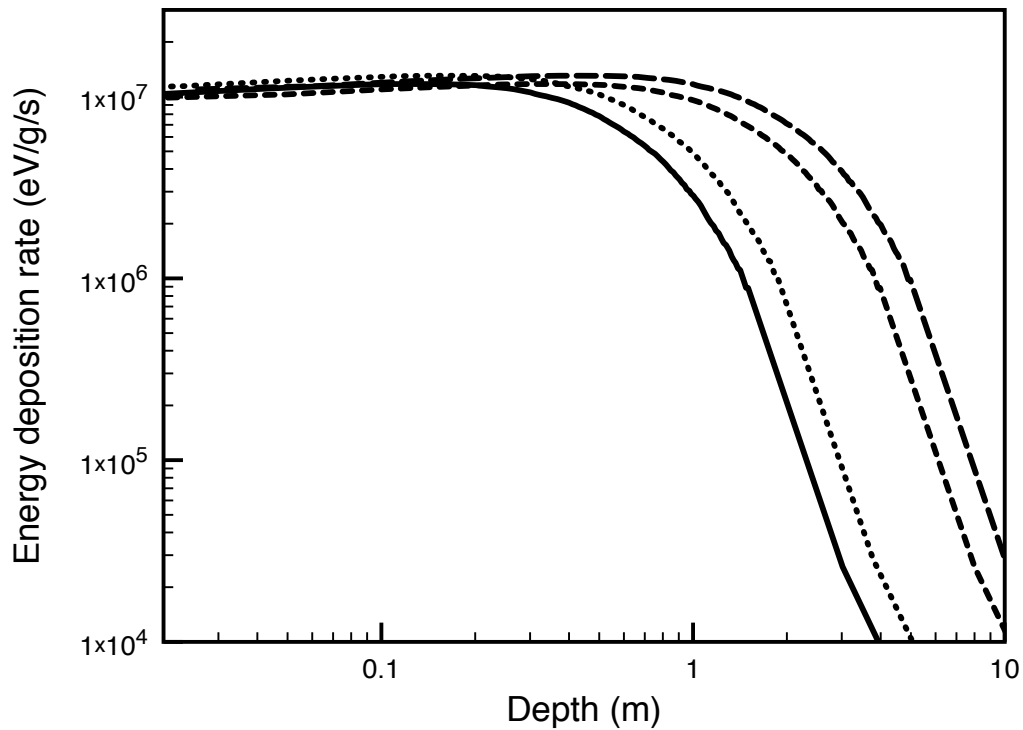

Figure 2: Simulation results of the subsurface energy deposition rate in different cases. Vertical axis is the energy deposition rate in units of eV/g/s and the horizontal axis is the depth in meters. Long dashes represent energy deposition in ice of a planet with no atmosphere (Europa/Enceladus), short dashes represent deposition for a planet with no atmosphere in rock with density 2.65 g/cm³ (Pluto), dots represent deposition in ice for Mars and solid line is for deposition in rock on Mars.

Three cases were considered here: a planetary object with no atmosphere so the entire GCR flux deposits energy in the ground; the Earth where most of the radiation is blocked by the atmosphere, and Mars. For each case the energy deposition in pure ice with 1 g/cm³ density and in standard rock with the density of 2.65 g/cm³ was calculated. The flux of GCRs was taken from Dorman (2004) and was incident directly on the planet's surface in the first case and on the planet's atmosphere in subsequent cases. The surface in case 1 is pure ice with 1 g/cm³ density and in case 2 is standard rock with density of 2.65 g/cm³. It must be noted that both pure ice and water give the same results in these simulations. $10^9$ particles were used for simulations and the energy deposition profile was obtained in each case. It must be emphasized that energy deposition calculations depend primarily on the column density of the material through which particles traverse and other factors such as temperature, pressure and atmospheric composition are not important. Figures 2 and 3 show the subsurface energy deposition rates for different cases. In Figure 2, the energy deposition rate is ~ $10^7$ eVg$^{-1}$s$^{-1}$ close to the surface and drops below $10^5$ eVg$^{-1}$s$^{-1}$ below in the 10m depth range. The deposition rate falls sharply in case of rock due to higher density compared to ice. Since the energy deposition rate is about three orders of magnitude lower in case of the Earth, it has been displayed separately in Figure 3. The atmosphere absorbs most of the radiation, however because of the long range of muons, a small amount of radiation reaches a depth of a few kms.

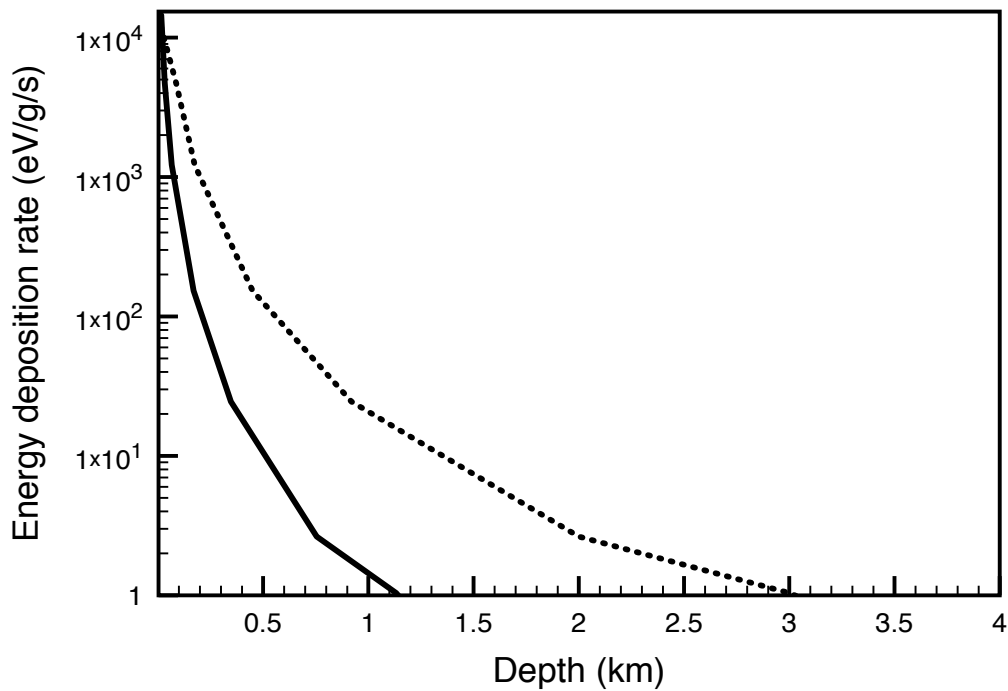

Figure 3: Subsurface energy deposition rate as a function of depth in rock of density 2.65 g/cm$^3$ (solid) and water/ice (dots). Vertical axis has energy deposition rate in eV/g/s and horizontal axis has depth in km.

Let us now compare these results with the energy environment of *Ca.* D. audaxviator. The radiolytic model calculations by Lin et al. (2005) yielded the net dosage range of alpha particles between $4.25 \times 10^5$ - $8.52 \times 10^5$ eVg$^{-1}$s$^{-1}$, beta particles between $6.58 \times 10^4$ - $4.27 \times 10^5$ eVg$^{-1}$s$^{-1}$, and for gamma rays between $4.00 \times 10^4$ - $2.25 \times 10^5$ eVg$^{-1}$s$^{-1}$ (Lin et al., 2005). This energy is used to split water molecules into H$^+$ and OH$^-$ forming H$_2$O$_2$. H$_2$O$_2$ in turn reacts with the surrounding medium to form sulfate, SO$_4^{2-}$. It uses sulfate instead of O$_2$ and obtains nitrogen from surrounding ammonia (Lin et al., 2006; Chivian et al., 2008). As one can see, the energy deposition rate is about an order of magnitude higher close to the surface in case 1 than that available to *Ca.* D. audaxviator.

Life could evolve a variety of mechanisms to utilize such a large range of energy injected underground. It could produce molecular hydrogen and oxidants useful for life. Muons can both directly react with molecules present in the medium, and also indirectly through radiolysis products (Smilga and Belousov, 1994). A detailed description of all the chemical reactions including the intermediate steps can be found elsewhere (Hatano et al., 2010). Let us now use these radiation doses to calculate familiar biologically useful processes such as the production of ATP. The terminal phosphate bond in ATP requires 0.304 eV per molecule (Schulze-Makuch and Irwin, 2008). As in case of *Ca.* D. audaxviator, the total energy produced by radioactive rocks would be potentially divided between radiolysis-powered metabolism, radiation damage and damage repair, only a fraction of the total energy will be utilized for radiolysis. Based on our simulation results the energy availability shown in Figure 2 ranges between ~$10^7$ to ~$10^4$ eVg$^{-1}$s$^{-1}$, which could potentially produce ATP molecules, and the upper limit of production in case is ~ $3 \times 10^7$ ATP molecules g$^{-1}$s$^{-1}$ based on the 0.304 eV per molecule conversion factor and assuming 100% efficiency.

It must be mentioned that life, as we know it, requires water, which is a neutral fluid and fits perfectly with temperature variations on Earth. However, other fluids might offer similar functionality in terms of being stable; they may provide transportation of essential nutrients and remain liquid for temperature ranges for that particular planet. Underground water or other fluid sources, in combination with flux of secondary particles can provide a stable self-sustained environment for life to exist. Low energy availability can produce organisms with a very slow metabolism. There is a possibility of an ecosystem thriving on this energy source based on other biochemical bases, and might necessitate alternate approaches to detect life 'as we don't know it' (Schulze-Makuch and Irwin, 2008; Azua-Bustos and Vega-Martínez, 2013) in subsurface environments on Earth and elsewhere.

A laboratory experiment to test this hypothesis could also be performed. It would involve gradually changing the radiation environment of *Ca.* D. audaxviator, by using different particle radiation beams, keeping the same chemical environment and observing its growth over a period of time. If the organism is able to adapt to gradually changing radiation environment, it can be eventually exposed to GCR secondaries proving an ultimate test for the hypothesis.

**2.2 GCR-induced synthesis of organics and other biologically useful products**

Even though *Ca.* D. audaxviator is able to maintain an ecosystem independent of photosynthesis, it uses photosynthetic products. Based on experimental results and theoretical models, some authors have proposed that high-energy particles could produce the amino acid glycine on extraterrestrial ices (Holtom et al., 2005). In this section, it is proposed that biologically useful products, using a similar mechanism can be produced in subsurface environments by GCR-induced secondaries. Charged particles directly interact with ice and produce a number of biologically useful secondary products (Bernstein et al., 1995; Kobayashi et al., 1995; Hudson and Moore, 1999). Hudson and Moore irradiated different mixtures of water and CO (carbon monoxide) with 0.8 MeV protons at temperatures near 16 K in order to simulate interstellar conditions. The results of isotopic substitution and IR spectroscopy showed the formation of several hydrocarbons such as HCOOH, HCO$^-$, H$_2$CO and CH$_3$OH (Hudson and Moore, 1999). Earlier experiments of Bernstein et al. were conducted with a larger temperature range (12 to 300 K), and in addition to the above mentioned products, they discovered hexamethylenetetramine (HMT, C$_6$H$_{12}$N$_4$), ethers, alcohols, compounds related to polyoxymethylene, ketones and amides in their samples (Bernstein et al., 1995). Their subsequent experiments showed the formation of aroma-bearing ketones and carbolylic acid functional groups (Bernstein et al., 2003). Other groups have also reported experimental evidence of the formation of amino acid precursors on exposure to high-energy particles (Kobayashi et al., 2000; Kobayashi et al., 2001). Kobayashi et al. (2000, 2001) irradiated several ice mixtures composed of methane, CO and ammonia with high-energy protons. The results of quadrupole mass spectrometry and ion exchange chromatography showed the formation of amino acids, such as glycine and alanine, and some hydrocarbons. Garrod and Herbst conducted charged particle induced photodissociation calculations to model chemical changes from interstellar radiation field and GCRs and reported the production of complex chemicals such as formic acid, methyl formate and dimethyl ether (Garrod and Herbst, 2006).

One common feature of these studies is that they all consider organic synthesis on the surface, which is true for high-energy photons such as UV, X-rays and low energy protons (~ keV – MeV). However for higher energy particles such as GCRs whose energies are ~ 10 GeV and beyond, the secondary particles penetrate below the surface. Particles such as electrons, positrons, neutrons and photons produced in interactions have very short ranges and are confined in a relatively small volume. Particles with the highest range are muons as shown in Figure 2 and are the primary source of GCR-induced radiation in such environments. In order to validate this hypothesis, one can irradiate samples with muons produced in accelerator experiments. Muons can also be approximated with electrons in laboratory experiments to a certain extent because electrons also undergo electromagnetic interactions like muons, however they lose energy very quickly, which makes this test possible only at low energies. Based on the studies cited above, additional mechanism supporting subsurface life could be direct organic synthesis induced by GCR-induced secondary particles, especially muons at greater depths. There is experimental evidence of the formation of amino acid precursors on exposure to high-energy particles (Kobayashi et al., 2000; Kobayashi et al., 2001). This mechanism could be especially important in case of comets, as cosmic ray induced ionization is believed to be the main driver of cometary organic chemistry (Cottin et al., 2000). Organic synthesis occurring at the polar regions of the Moon, Mercury and other silicate bodies has also been proposed (Crites et al., 2013). There are studies of GCR-induced synthesis of organic molecules in Titan's atmosphere (Capone et al., 1980; Capone et al., 1981), production of oxidants on Europa (Chyba and Hand, 2001) and the possibility of an aerial biosphere on Venus (Dartnell et al., 2015).

The production of $O_2$, $H_2O_2$ and other oxidants on Europa's surface by charged particles accelerated in Jovian magnetic field has been estimated in an earlier study (Chyba and Hand, 2001). They proposed that through impact gardening, these biologically useful chemicals could be transported to Europa's oceans. Ionization through $^{40}K$ decay was also considered. Since GCRs are much more energetic compared to particles considered in the above study, in addition to producing these chemicals on Europa's surface, GCR secondaries can directly produce them in Ice below the surface. Let us now calculate the energy deposited in the ice shell of Europa from GCRs. The average energy deposited from 0 to 1 meter depth is ~ $10^{15}$ eV/g/yr. For this 1-m shell for entire Europa, the total energy is ~ $1.6 \times 10^{32}$ eV/yr, which would produce $2.4 \times 10^7$ moles/yr of $H_2$ and $O_2$. This scenario is valid in other cases too where non-photosynthetic chemicals can be produced in subsurface environments using this mechanism.

3. Implications on the origin of life and possibility of finding life beyond Earth

It is believed that ~ 3.5 - 4 Gyr ago, when life originated on the Earth, the sun was in a highly active phase. In this scenario, the Earth's surface was likely to be bombarded by a high flux of energetic solar particles and super CMEs - Coronal Mass Ejections. Enhanced flux of solar particles and variability in the GCR flux can also enhance the rate of lightning (Erlykin and Wolfendale, 2010), and provide energy to the prebiotic soup to synthesize amino acids and other organic compounds forming the building blocks of life (Miller et al., 1976). If the solar particles are sufficiently energetic (>10 GeV) they can produce secondary particles capable of penetrating underground and in water (Khalchukov et al., 1995; Atri and Melott, 2011), providing a small source of energy away from direct exposure to high flux of harmful UV radiation on the surface. Solar particles typically reach energies of 100s of MeV during violent eruptions and in some cases greater than 10 GeV (Tylka and Dietrich, 2009). Typical SPEs (Solar Proton Events) produce a fluence of about $10^9$ protons/cm$^2$ on Earth. Since the Sun was considerably more active in the past, it is highly likely that such eruptions might have occurred on the Sun more frequently. The higher energy component of these SPEs (Tylka and Dietrich, 2009), just like GCRs is capable of penetrating underground (Atri and Melott, 2011), and would have increased the flux of secondary particles in subsurface environments. A combination of particle flux along with water and nutrients might provide ideal conditions for life to originate and evolve until conditions on the surface become optimal.

| Source | Energy availability |
|---|---|
| GCR Mars subsurface | $1.2 \times 10^7$ eV/g/s |
| GCR Europa, Enceladus, Pluto, Moon subsurface | $1.3 \times 10^7$ eV/g/s |
| GCR Earth subsurface | $1.5 \times 10^4$ eV/g/s |
| Radioactive rocks | $1.0 \times 10^6$ eV/g/s |
| Hydrothermal vents | $2.5 \times 10^6$ eV/°C |
| Geochemical | 1.3 eV/e$^-$ transfer |

Table 1: Energy availability from GCR simulations and other sources. GCR results show the maximum energy deposited, which is near the surface and drops with depth. For radioactive rocks, the value is the average obtained from Lin et al. (2005). For hydrothermal vents, the energy availability is obtained from Schulze-Makuch & Irwin (2008) and depends on temperature gradient. The Gibbs free energy of 145 reactions was calculated by Rogers and Amend (2005) and ranged from 0 to 125 kJ/mol (1.3 eV/e$^-$ transfer) per electron transfer. The value is the table shows the maximum energy from a reaction.

Since independent/freely-floating or "rogue" planets are not tied to any stellar system, they do not receive a steady stream of photons from a parent star. A mechanism has been proposed which could support life on such planets with a combination of sufficient pressure and radioactive heat (Stevenson, 1999). Alternatively, GCRs and radioactive materials can be a steady source of energy on such planets. The mechanisms proposed in this paper can be used to synthesize biologically useful chemicals and to power such ecosystems. Europa is believed to have an abundance of liquid water below its thick ice shell (Chyba and Phillips, 2001). GCR-induced particles, although cannot provide energy in the ocean, as discussed earlier, they can provide ingredients and fuel to a potential ecosystem in its ice shell. Table 1 shows the energy availability from a number of sources on different planetary objects. Objects with no or negligible atmospheres have higher energy availability, about one order-of-magnitude higher than that available to *Ca.* D. audaxviator from radioactive rocks. There are no other such organisms found on Earth so far. One reason could be that due to a substantial atmosphere, most of the radiation is blocked and only a small amount is available in subsurface environments. As seen in Table 1, the energy availability on Earth is three orders-of-magnitude smaller than that found on Mars and about two orders-of-magnitude smaller than used by *Ca.* D. audaxviator. As in case of *Ca.* D. audaxviator, where the total energy produced by radioactive rocks is divided between radiolysis-powered metabolism, radiation damage and damage repair, it is possible that a fraction of this energy deposited by GCRs can be utilized for metabolic purposes.

There is growing evidence of pockets of near-surface water on Mars (Chojnacki et al, 2016). The presence of indigenous nitrogen in sedimentary and aeolian deposits using the SAM instrument on-board the Curiosity rover was reported recently (Stern et al., 2015). Observations of hydrated salts – magnesium perchlorate, magnesium chlorate and sodium perchlorate on the Martian surface were also recently announced (Ojha et al., 2015). Radiolysis-powered ecosystems can use these chemicals for metabolic processes. The possibility of methane, $N_2$ and traces of $O_2$ being by-products of such an ecosystem cannot be ruled out and could possibly explain the presence of methane (Formisano et al., 2004; Webster et al., 2015) that cannot yet be explained by standard physics and chemistry models (Atreya et al., 2007; Lefevre and Forget, 2009).

## 4. Conclusions

Studying the biological effects of ionizing radiation is a growing area of research. Much of the effort has been focused on examining its damaging effects on human health in context of radiation oncology, nuclear accidents and astronaut health in outer space. Several experiments have shown ionizing radiation to synthesize organic compounds on interaction with ice mixtures. The discovery of *Ca.* D. audaxviator thriving 2.8 km below the Earth's surface powered by radiolysis opens up new possibilities of biological interaction with (ionizing) particle radiation. GCRs produce secondary particles that deposit energy in the subsurface environment. Conceivable mechanisms have been proposed through which the energy of GCR-induced particles can be used to produce biologically useful products such as organics and utilize energy for radiolysis to power a potential subsurface ecosystem. Ionizing radiation causes damage, and just as in case of *Ca.* D. audaxviator, a part of the energy deposited in subsurface environments can be used for repairing damage and the rest for chemical reactions and potential biological use.

GCRs are a steady source of ionizing radiation throughout the Galaxy and beyond. Their secondary component can deposit energy underground; muons especially can penetrate several kilometers underground (Atri and Melott, 2011; Atri and Melott, 2014; Dorman, 2004; Gaisser, 1990; Groom et al., 2001; Khalchukov et al., 1995; Mei and Hime, 2006). It has been shown that GCR-induced radiolysis is a steady source of energy for subsurface environments and could potentially be a viable source of energy supporting such an ecosystem. GCR-induced particles can directly interact with the medium with essential nutrients and synthesize basic chemicals vital for life to develop, analogous to the experiments with high-energy protons and ice mixtures (Cottin et al., 1999; Garrot and Herbst, 2006; Holtom et al., 2005; Hudson and Moore, 1999; Kobayashi et al., 1995; Kobayashi et al., 2001).

The GCR-induced radiolysis mechanisms proposed in the paper open up new possibilities of life in subsurface environments on a number of planetary bodies such as Mars, Moon, Europa, Enceladus, Pluto, and especially ones with negligible atmospheres. Radiolysis-powered life can either thrive independently, or can consume a combination of sources such as heat from chemical and geological processes. GCR-induced radiolysis can produce a number of ions species leading to the production of biological useful products such as molecular hydrogen. There is a possibility of life on icy objects in the interplanetary medium such as comets, and other bodies in the interstellar environment. This energy source could support life locked inside icy objects and facilitate efficient transportation conferring to the panspermia hypothesis. Since rogue or independent planets also receive a steady flux of this radiation, there is a possibility of a thriving subsurface ecosystem on such planets. Ground-based laboratory tests are suggested that can be conducted to validate the hypotheses presented here.

**Acknowledgements**

The author acknowledges Arnold Wolfendale, Andrew Karam, Adrian Melott, Jacob Haqq-Misra, Steven Hsu and the anonymous reviewers for their helpful comments, and Dipanwita Shome for her help with editing the manuscript. This work made use of the Extreme Science and Engineering Discovery Environment, supported by the National Science Foundation grant number ACI-1053575.

**List of abbreviations:**

ATP - adenosine triphosphate
CMEs - Coronal Mass Ejections
CO – Carbon Monoxide
DNA - deoxyribonucleic acid
eV – electron volt
GCR – Galactic Cosmic Rays
Gy – Gray
LD – Lethal Dose
SPEs - Solar Proton Events
UV – Ultra Violet